\documentclass[12pt]{article}
\usepackage{psfig}
\usepackage{times}

\def\go{\mathrel{\raise.3ex\hbox{$>$}\mkern-14mu\lower0.6ex\hbox{$\sim$}}}
\def\lo{\mathrel{\raise.3ex\hbox{$<$}\mkern-14mu\lower0.6ex\hbox{$\sim$}}}

\title{Planet--Planet Scattering in Upsilon Andromedae}
\author
{
Eric B.\ Ford,$^1$ Verene Lystad,$^{2}$ Frederic A.\ Rasio$^{2}$\\
\\
\normalsize{$^{1}$Department of Astronomy, University of California,
Berkeley, CA, USA}\\
\normalsize{$^{2}$Department of Physics and Astronomy, Northwestern
University, Evanston, IL, USA}\\
}

\begin{document}



\maketitle

\bigskip
\centerline{To appear in {\it Nature\/}, March 2005}
\bigskip
\bigskip

{\bf 
Doppler spectroscopy has detected 136 planets around nearby stars$^1$.
A major puzzle is why their orbits are highly eccentric, while all
planets in our Solar System are on nearly circular orbits, as expected
if they formed by accretion processes in a protostellar disk. Several
mechanisms have been proposed to generate large eccentricities after
planet formation, but so far there has been little observational
evidence to support any particular one.  Here we report that the
current orbital configuration of the three giant planets around
Upsilon Andromedae$^{2,3}$ ($\upsilon$ And) provides evidence for a
close dynamical interaction with another planet$^4$, now lost from the
system.  The planets started on nearly circular orbits, but chaotic
evolution caused the outer planet ($\upsilon$ And~d) to be perturbed
suddenly into a higher-eccentricity orbit.  The coupled evolution of
the system then causes slow periodic variations in the eccentricity of
the middle planet ($\upsilon$ And~c).  Indeed, we show that $\upsilon$
And~c periodically returns to a very nearly circular state every 9000
years.  Our analysis shows that strong planet--planet scattering, one
of several mechanisms previously discussed for increasing orbital
eccentricities, must have operated in this system.
}


The innermost planet around $\upsilon$ And has a very small orbital
eccentricity, as expected from tidal circularization$^{5}$, but the
two outer planets have large eccentricities, both around 0.3. It was
quickly recognized after their discovery that the gravitational
interaction between these two planets causes significant eccentricity
evolution on secular timescales ($\sim10^4$ years).  In particular, in
some early solutions, the middle planet appeared to have its
eccentricity varying periodically with a large amplitude, from a
maximum near the present value to a minimum near zero$^{6}$. Later
fits to the data suggested that the outer two orbits had arguments of
pericenter nearly equal to within about $10^\circ$.  This could
indicate an ``apsidal resonance,'' in which two elliptical orbits
oscillate about an aligned configuration with a small libration
amplitude$^{7}$.

Two possible explanations have been proposed for this peculiar orbital
configuration. The first is a dynamical mechanism in which a sudden
perturbation imparts a finite eccentricity to the outer
planet$^4$. The subsequent secular evolution causes the middle
planet's orbit to become eccentric and can leave the two orbits either
circulating, or librating with a large amplitude.  In either case, the
eccentricity of the middle planet periodically returns to its initial
low value.  The impulsive perturbation of the outer planet would
result naturally from planet--planet scattering, which was one of the
earliest mechanisms proposed for inducing large eccentricities in
extrasolar planets$^{8,9}$.  In contrast, the second explanation
invokes an adiabatic perturbation of the outer planet's eccentricity
through torques exerted by an exterior gaseous disk$^{10}$.  As the
eccentricity of the middle planet grows on a similarly long timescale,
this would leave the system in an apsidal resonance by damping the
libration amplitude to zero.  This is a natural extension of more
general ``migration scenarios'' in which the coupling of a planet's
orbit to a gaseous disk could both increase eccentricities$^{11}$ and
lead to orbital decay and the formation of planets with very short
orbital periods$^{12}$.

The $\upsilon$ And system was the first extrasolar multi-planet system
ever discovered by Doppler spectroscopy.  Since the first announcement
of the outer two planets in 1999, the California and Carnegie Planet
Search team has taken over 350 new radial velocity measurements,
making $\upsilon$ And one of the systems with the tightest constraints
on orbital parameters (Table~1; see also supplementary information).
The secular evolution of the system is shown in Fig.~1.  While the
eccentricity of~d remains always between 0.2 and 0.3, planet~c returns
periodically to a nearly circular orbit with $e<0.01$.  As shown in
Fig.~2, this is now a property of {\it all\/} solutions consistent
with the data, in contrast to earlier suggestions that it could happen
for {\it some\/} solutions$^{4,6}$.

We have also re-examined the possible presence of apsidal resonance in
the system.  Remarkably, we find that the allowed solutions all lie
very close to the boundary between librating and circulating
configurations (Fig.~3). As a consequence, all librating systems have
libration amplitudes close to $90^\circ$.  As shown in Fig.~3, this
behavior is confirmed by direct numerical integrations of all three
planets for a large sample of systems covering the allowed parameter
space of solutions.

Our results do not change significantly if the assumption of a
coplanar system viewed edge-on is relaxed. Inclination angles can be
constrained by requiring dynamical stability of the
system$^{6,7,16,17}$.  Using the latest data we find that, for a
coplanar configuration of three planets with the best-fit orbital
parameters, the system becomes dynamically unstable when $\sin i<0.5$
(where $\sin i=1$ corresponds to edge-on). If we relax coplanarity, we
find that the system becomes dynamically unstable whenever the
relative inclination is greater than about $40^\circ$.  Throughout the
full range of allowed inclination angles we find qualitatively similar
behavior for the secular evolution of the system. For example, Fig.~2
shows that the eccentricity of planet~c would still return
periodically to a value near zero.  The secular evolution of the
system should be reevaluated if future observations of $\upsilon$ And
were to discover an additional planet in a long period orbit.

Our analysis clearly confirms that the $\upsilon$ And system is
evolving exactly as would be expected after an impulsive perturbation
to $\upsilon$ And~d$^4$.  The initial sudden change in eccentricity
for the outer planet would be naturally produced by a close encounter
with another planet, which got ejected from the system as a
result. Using our knowledge of planet--planet scattering from several
previous studies$^{2,18-20}$, we determined plausible initial
conditions for the original, unstable system. The early dynamical
evolution of such a system is illustrated in Fig.~4.  After a brief
period of strongly chaotic evolution, lasting $\sim10^3$ years, the
outer planet is ejected, and the remaining two planets are left in a
dynamical configuration closely resembling that of $\upsilon$ And (see supplementary
discussion for a more detailed discussion).

While several other mechanisms (e.g., perturbations in a binary
star$^{22}$, resonances$^{10}$, interaction with a gaseous
disk$^{11}$) have been proposed to explain the large eccentricities of
extrasolar planets, only planet--planet scattering naturally results
in an impulsive perturbation, as is necessary to explain the
$\upsilon$ And system. All other mechanisms operate on much longer
timescales and would also affect the eccentricity of planet~c, erasing
the memory of its initial circular orbit (see supplementary
discussion for a more detailed discussion).

Our results have other implications for planet formation.  Given the
difficulty of forming giant planets at small orbital distances, it is
generally assumed that the $\upsilon$ And planets migrated inward to
their current locations via interactions with the protoplanetary
disk$^{12}$.  If this is correct, then the small minimum eccentricity
of $\upsilon$ And~c also provides evidence that its eccentricity at
the end of migration had not grown significantly, in contrast to what
some theories predict$^{11}$.  However, the possibility of formation
{\it in situ\/}$^{23}$ cannot be excluded by our results.

\bigskip

\hrule
\medskip

\begin{enumerate}
\item Schneider, J. Extra-solar planets catalog.
(http://cfa-www.harvard.edu/planets/catalog.html)
(2004).
\item Butler, R.P. {\it et al.\/} Evidence for multiple companions to
Upsilon Andromedae.
{\it Astrophys.\ J.\/} {\bf 526}, 916--927 (1999).
\item Fischer, D.A. {\it et al.\/} A planetary companion to HD 40979
and additional planets orbiting
HD 12661 and HD 38529, {\it Astrophys.\ J.\/} {\bf 586}, 1394--1408
(2003).
\item Malhotra, R. A dynamical mechanism for establishing apsidal
resonance.
   {\it Astrophys.\ J.\ Letters\/} {\bf 575}, L33--L36 (2002).
\item Rasio, F.A., Tout, C.A., Lubow, S.H. \& Livio, M. Tidal decay of
close
planetary orbits.
   {\it Astrophys.\ J.\/} {\bf 470}, 1187--1191 (1996).
\item  Stepinski, T.F., Malhotra, R. \& Black, D.C. The Upsilon
Andromedae system:
models and stability.
{\it Astrophys.\ J.\/} {\bf 545}, 1044--1057 (2000).
\item Chiang, E.I., Tabachnik, S. \& Tremaine, S. Apsidal alignment in
Upsilon Andromedae.
{\it Astron.\ J.\/} {\bf 122}, 1607--1615 (2001).
\item Rasio, F.A. \& Ford, E.B. Dynamical instabilities and the
formation of extrasolar planetary systems. {\it Science\/} {\bf 274},
954--956 (1996).
\item Weidenschilling, S.J. \& Marzari, F. Gravitational scattering as
a possible origin for giant planets at small stellar distances. {\it
Nature\/} {\bf 384}, 619--621 (1996).
\item Chiang, E.I. \& Murray, N. Eccentricity excitation and apsidal
resonance capture
in the planetary system Upsilon Andromedae.
{\it Astrophys.\ J.\/} {\bf 576}, 473--477 (2002).
\item Goldreich, P. \& Sari, R. Eccentricity evolution for planets in
gaseous disks.
   {\it Astrophys.\ J.\/} {\bf 585}, 1024--1037 (2003).
\item Lin, D.N.C. {\it et al.\/} in {\it Protostars and Planets IV\/}
(eds Mannings, V., Boss, A.P. \& Russell, S.S.) 1111--1178
(University of Arizona Press, Tucson, 2000).
\item Ford, E.B. Quantifying the uncertainty in the orbits of
extrasolar planets.
  {\it Astron.\ J.\/}, in press (2005).
\item Chambers, J.E. A hybrid symplectic integrator that permits close
encounters between massive bodies
{\it Mon.\ Not.\ R.\ Astron.\ Soc.\/} {\bf 304}, 793--799 (1999).
\item Murray, C.D. \& Dermott, S.F. {\it Solar System Dynamics\/}
(Cambridge University Press, New York, 1999).
\item Lissauer, J.J. \& Rivera, E.J. Stability analysis of the
planetary system orbiting Upsilon Andromedae. II. Simulations using new
Lick Observatory
fits.
   {\it Astrophys.\ J.\/} {\bf
554}, 1141--1150 (2001).
\item Lystad, V. \& Rasio, F. in {\it The Search for Other Worlds\/}
(eds Holt, S.S. \& Deming, D.) 273--276
(AIP Conf.\ Proc.\/ 713, American Institute of Physics, 2004).
\item Ford, E.B., Havlickova, M. \& Rasio, F.A. Dynamical instabilities
in extrasolar planetary systems containing two giant planets. {\it
Icarus\/} {\bf 150}, 303--313 (2001).
\item Ford, E.B., Rasio, F.A. \& Yu, K. in {\it Scientific Frontiers
in Research on Extrasolar Planets\/}
(eds Deming, D. \& Seager, S.) 181--187 (ASP Conf.\ Ser.\ 294,
Astronomical Society of the Pacific, San Francisco, 2003).
\item Marzari, E.F. \& Weidenschilling, S.J.  Eccentric extrasolar
planets: the
jumping jupiter model. {\it Icarus\/} {\bf 156}, 570--579  (2002).
\item Gladman, B. Dynamics of systems of two close planets.
{\it Icarus\/} {\bf 106}, 247--265 (1993).
\item Holman, M., Touma, T. \& Tremaine, S.
Chaotic variations in the eccentricity of the planet orbiting 16 Cyg B.
{\it Nature\/} {\bf 386}, 254--256 (1997).
\item Bodenheimer, P., Hubickyj, O. \& Lissauer, J.J.
Models of the in situ formation of detected extrasolar giant planets.
{\it Icarus\/} {\bf 143}, 2--14 (2000).
\end{enumerate}
\medskip

\noindent
Supplementary information accompanies the paper on
www.nature.com/nature.
\medskip

\noindent
{\bf Correspondence} and requests for materials should be addressed to
F.A.R. (rasio@northwestern.edu).
\medskip

\noindent
{\bf Acknowledgments} We are very grateful to Debra Fischer for
providing us with the
latest radial velocity data on $\upsilon$ And.  We also thank
Eugene Chiang, Man Hoi Lee and Stan Peale for useful discussions. This
work was supported by an NSF Grant to FAR at
Northwestern University and by a Miller Research Fellowship to EBF. VL
acknowledges support from the NASA Undergraduate Summer Research
Program at Northwestern. FAR and EBF thank the Kavli
Institute for Theoretical Physics for hospitality and support.
\medskip

\noindent
{\bf Competing interests statement} The authors declare that they have
no competing financial interests.

\clearpage

\begin{table}
\center{\bf Masses and orbital parameters for the three planets in
$\upsilon$ And.}
\begin{center}
\begin{tabular}{lllll}
\hline
    & & & & \\
Planet~~~~ & $P$ (d)      & $e$       & $\omega$ (deg) & $m \sin i$
($M_{\mathrm J}$) \\
   & & & & \\
b      & 4.617146(56)~~~~ & 0.016(11)~~~~ &                & 0.6777(79)
\\
\\
c      & 241.32(18)       & 0.258(15)     & 250.2(4.0)~~~~ & 1.943(35)
\\
\\
d      & 1301.0(7.0)      & 0.279(22)     & 287.9(4.8)     & 3.943(57)
\\
    & & & & \\
\hline
\end{tabular}
\end{center}
\caption{Results of our new analysis of the $\upsilon$ And radial
velocity data$^{2,3,13}$. We have used the entire Lick
Observatory data set, kindly provided to us by D.~Fischer.   For
conciseness, we present only the means and standard deviations on the
last two digits (indicated in parenthesis) after marginalizing over
all other parameters.  We list the orbital period ($P$) in days, the
orbital eccentricity ($e$), the argument of pericenter ($\omega$) in
degrees, and the planet mass times the sine of the inclination of the
orbital plane to the line of sight ($m \sin i$) in units of Jupiter
masses ($M_{\rm J}$). More details on our analysis are presented in
the supplementary information.
}
\end{table}

\clearpage

\begin{figure}
\centerline{\psfig{file=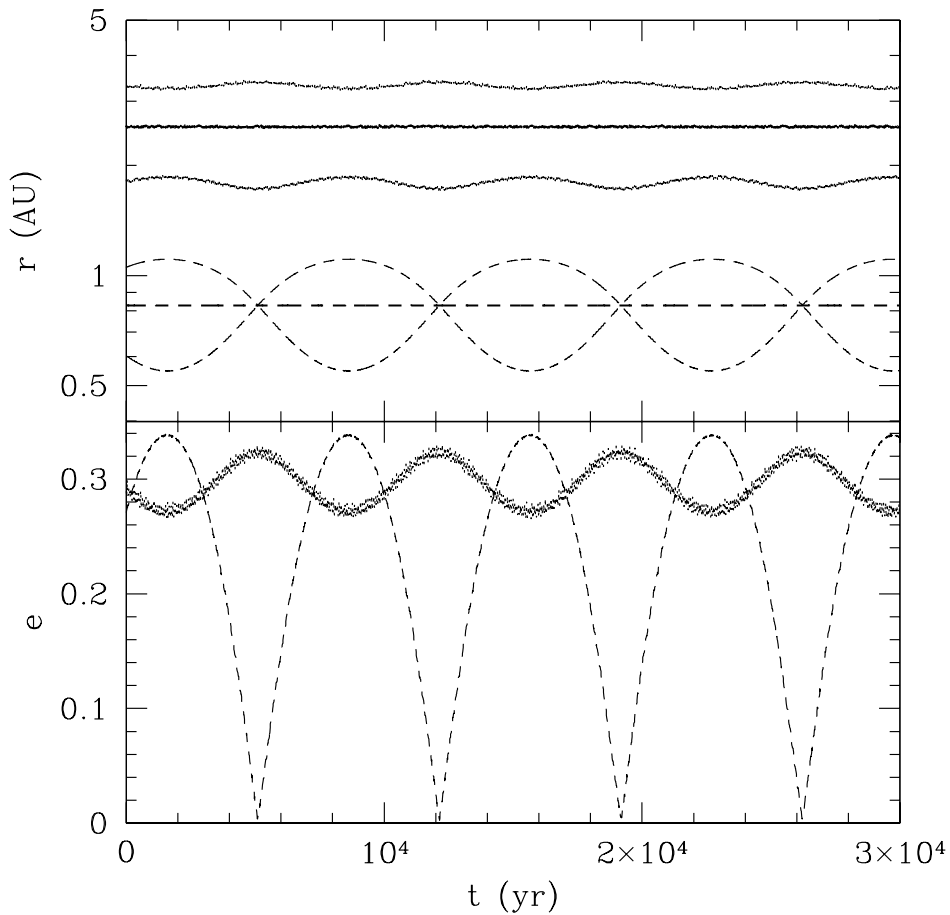,width=13cm}}
\caption{
Secular evolution of the planetary system around $\upsilon$
And. The top panel
shows the semimajor axes (thick lines), as well as the periastron and
apastron distances (thin lines), for the outer two planets.  The
lower panel shows the evolution of the orbital eccentricity for each
planet.  Note that both planet~c (dashed) and planet~d (dotted) have a
significant eccentricity at the present time ($t=0$), but that the
eccentricity of~c returns periodically to very small values near zero.
The results shown here were obtained by direct numerical integration
using our best-fit parameters.
All direct $N$-body integrations presented in this paper
were performed using {\tt Mercury}$^{14}$, version~6.1.
}
\end{figure}

\begin{figure}
\centerline{\psfig{file=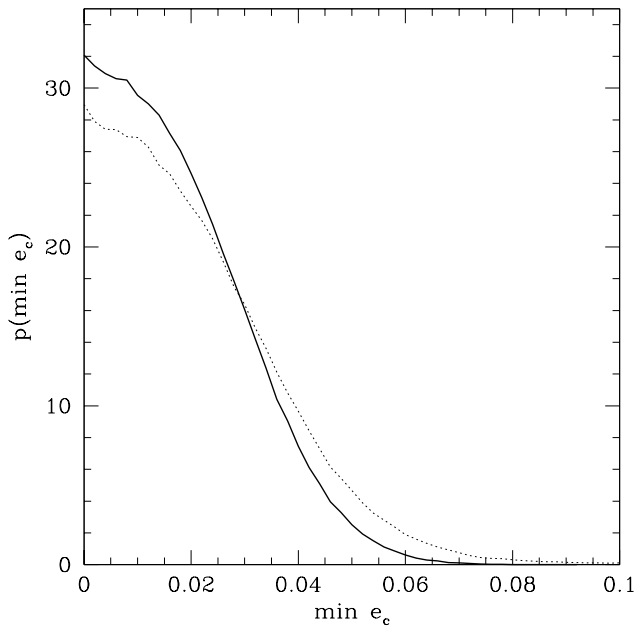,width=9cm}}
\caption{ Probability distribution for the minimum eccentricity of
planet~c.  We draw initial orbital elements for  planets~b,~c,
and~d from their posterior probability distribution and evolve each
system
according to classical second-order secular perturbation theory. We
find that {\it all\/} allowed orbital solutions have the eccentricity
of planet~c oscillating from a maximum value slightly larger than the
present value to very nearly zero. The solid curve
corresponds to coplanar orbits viewed edge-on. The dotted
curve shows the result for orbits with random orientations
to the line of sight, but requiring dynamical stability. This implies
relative inclinations $< 40^\circ$.
Note that systems with relative inclinations  $> 140^\circ$
are also dynamically stable.  Although such retrograde orbits are
unlikely
on theoretical grounds, our conclusions are robust to this
possibility.  Since the secular perturbation theory averages over the
orbits, it is also valid for retrograde orbits. 
The fact that all allowed solutions result in the eccentricity of
planet~c returning to a very small value can be understood easily from
lowest-order secular perturbation theory,$^{4,15}$ where the
eccentricity vector of each planet can be described as the sum of 
three rotating eigenvectors in the $(e\cos\omega, e\sin\omega)$
plane. The eigenvector representing the effects of planet~b on
planet~c has a very small amplitude and can be neglected.  For the
particular configuration of $\upsilon$ And, the two dominant
eigenvectors describing planet~c have very nearly the same
length. Depending on whether the eigenvector with the faster rotation
(higher eigenfrequency) has a slightly larger or smaller length than
the other, the vector sum will be rotating around $360^\circ$
(circulation) or oscillating with an amplitude close to $90^\circ$
(libration), respectively. Whenever the two vectors are anti-aligned,
the magnitude of their sum, i.e., the eccentricity of the planet, is
very close to zero; when they are aligned, the eccentricity is
maximum. 
}
\end{figure}

\begin{figure}
\centerline{\psfig{file=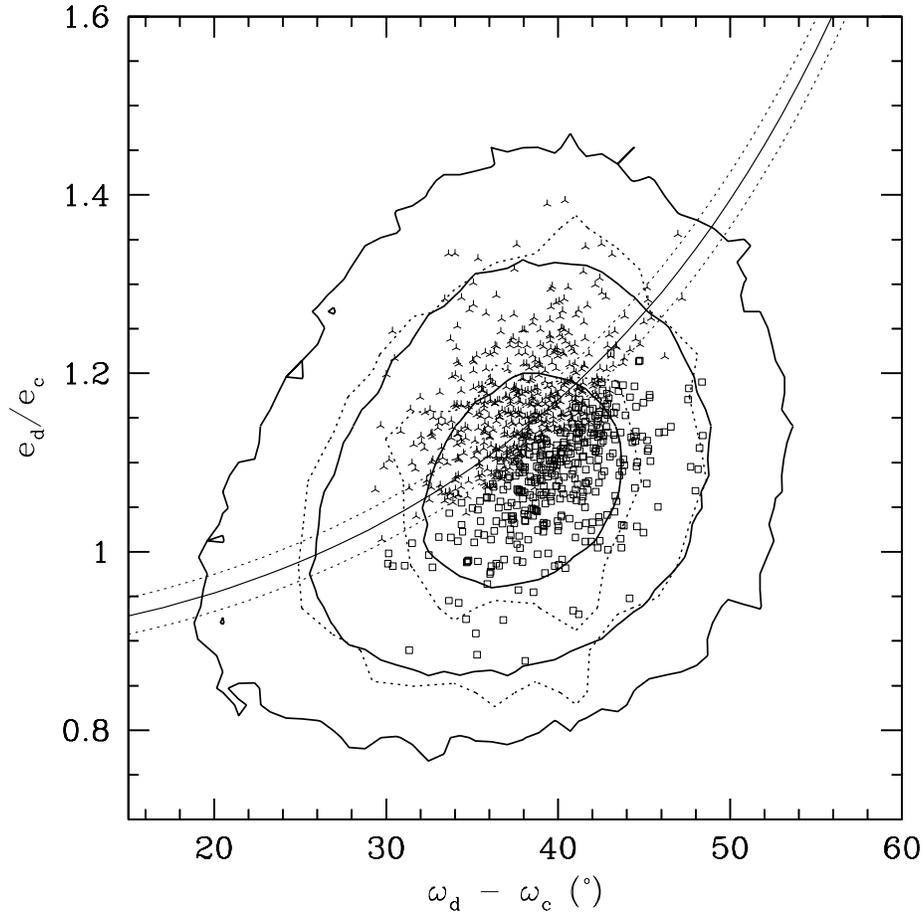,width=13cm}}
\caption{ Observational constraints on the secular evolution
parameters.  The eccentricity ratio is plotted as a function of the
difference between arguments of pericenter for planets~c and~d, all at
the present epoch.  We show the 1-, 2-, and 3-$\sigma$ contours for
the posterior probability distribution function marginalized over the
remaining fit parameters.  The thick contours assume that the
radial velocity variations are the result of three non-interacting
Keplerian orbits viewed edge-on, while the dotted contours
include the mutual gravitational interactions of the planets when
fitting to the radial velocity data (only 1- and 2-$\sigma$ contours
are shown).  The thin
solid line shows the separatrix between the librating (upper left) and
circulating (lower right) solutions according to the classical
second-order perturbation theory (neglecting the inner planet~b).  The
dotted lines on either side show the variation in the location
of the separatrix due to the uncertainty in the remaining orbital
elements.  The data points show the results of our direct numerical
integrations for the full three-planet system: triangles (squares)
indicate that the system was found to be librating (circulating).
Note that, regardless of the assumptions, the separatrix runs right
through the 1-$\sigma$ contours.  We find the median libration amplitude
of the librating systems to be about $80^\circ$. }
\end{figure}

\begin{figure}
\centerline{\psfig{file=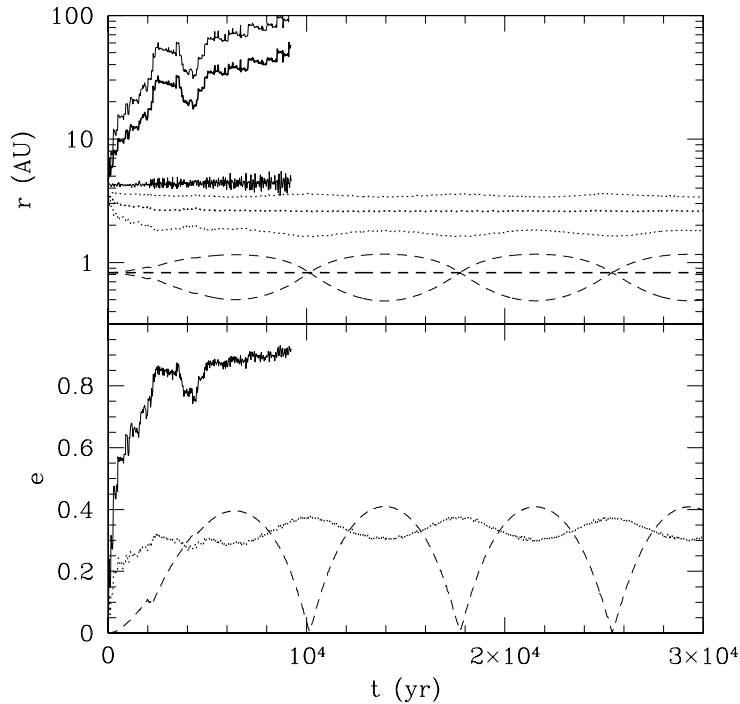,width=10cm}}
\caption{ Dynamical evolution of a hypothetical planetary system
similar to $\upsilon$ And.  After a brief period of dynamical
instability, one planet is ejected, leaving the other two in a
configuration very similar to that of $\upsilon$ And~c and~d.  The
innermost planet was not included, as it plays a negligible role.  All
planets were initially on nearly circular orbits. The initial periods
of the two inner planets were 241.5 days (equal to the period of
$\upsilon$ And~c) and 2100 days.  An additional planet (solid), of
mass $1.9\,M_{\rm J}$, was placed on an orbit of period 3333 days, so
that the outer two planets were close to their dynamical stability
limit$^{21}$.  The fourth planet (solid) interacted strongly with the
third (dotted), while the second planet (dashed) maintained a small
eccentricity. After the ejection, the two remaining planets evolve
secularly just like $\upsilon$ And~c and~d (cf.\ Fig.~1).  While this simulation
illustrates a case in which $\upsilon$ And had only one extra planet,
our results do not preclude the existence of even more planets at
larger distances.  In fact, the presence of another planet in an even
longer-period orbit could be responsible for triggering the
instability after a long period of seemingly ``stable'' evolution  
and help raise more quickly the pericenter of the planet which
perturbed $\upsilon$ And~d.  Note that the timescale to completely
eject the outer planet from the system (after $\sim9000$ years in this
particular simulation) is much longer than the timescale of the
initial strong scattering. After this initial brief phase of strong
interaction, the perturbations on the outer planet are too weak to
affect significantly the coupled secular evolution of $\upsilon$ And c
and d. Thus, the ``initial'' eccentricity of $\upsilon$ And c for the
secular evolution is determined by its value at the end of the strong
interaction phase, rather that at the time of the final ejection.  }
\end{figure}

\clearpage

\begin{center}
\noindent {\bf Supplementary Methods}
\end{center}

We model the observations as resulting from three planets on
independent Keplerian orbits. We use a Bayesian framework$^{s1}$, to
constrain the orbital elements and masses with the radial velocity
observations, as well as the ``stellar jitter'' (radial velocity
variations due to stellar phenomena such as convection, star spots,
and rotation).  We assume priors uniform in the logarithms of orbital
periods, velocity semi-amplitudes, and the stellar jitter.  Initially,
we assume that the orbits are coplanar and viewed edge-on.  (Later, we
relax this assumption by adopting instead an isotropic distribution of
inclinations but rejecting configurations that are not dynamically
stable.)  We assume uniform priors for the eccentricities and the
remaining angles.

We use the techniques of Markov chain Monte Carlo (MCMC) and the
Metropolis-Hastings algorithm with the Gibbs sampler to sample from
the posterior probability distribution for the masses and orbital
parameters$^{s2}$.  Since our Bayesian analysis closely follows the
methods developed for single-planet systems with no stellar jitter by
Ford$^{s2}$, here we only discuss the generalizations to the algorithm
which were necessary to apply it to multiple-planet systems and to
allow for an unknown stellar jitter.  Each state of the Markov chain
includes the five fit parameters for each planet (orbital period, $P$,
velocity semi-amplitude, $K$, orbital eccentricity, $e$, argument of
pericenter, $\omega$, and mean anomaly at the specified epoch, $M_o$),
the unperturbed stellar velocity, $C$, and the magnitude of the
stellar jitter, $\sigma_j$.  The jitter is assumed to be Gaussian and
uncorrelated from one observation to the next.  We use the Gibbs
sampler and Gaussian candidate transition functions which are centered
on the parameter values from the current state in the Markov chain.
The scales of the candidate transition functions were chosen based on
a preliminary Markov chain so as to result in acceptance rates of
nearly 40\%.  The computational efficiency of MCMC can also be
significantly affected by correlations between variables.  To improve
the computational efficiency we added candidate transition functions
which were based on several auxiliary variables based on combinations of
the fit parameters.  These auxiliary variables are $e \sin \omega$, $e
\cos \omega$, $\omega+M_o$, $\omega-M_o$, $P^{2/3} (1-e)$, and
$P^{2/3} (1+e)$, for each planet, as well as $\omega_b\pm\omega_c$,
and $\omega_c\pm\omega_d$ (where $\omega_b$, $\omega_c$ and $\omega_d$
are the arguments of pericenter for planets b, c, and d,
respectively).  The acceptance ratio was determined according to the
Metropolis-Hastings algorithm to reflect our choice of priors
described in the previous paragraph.  The use of this enlarged set of
candidate transition functions significantly improved the mixing and
hence efficiency of our Markov chains.  It is important to note that
our sampling procedure does not make any assumptions about the
posterior distribution for the orbital elements or masses. Therefore,
our approach allows us to take into account correlations between
various orbital parameters, in contrast to previous simpler analyzes.

We have computed five Markov chains, each of which contains over one
million states.  We have performed severals checks to verify the
reliability of our Markov chains.  The acceptance rates were between
0.37 and 0.45 for trail states generated by the candidate transition
functions for each variable or auxiliary variable in each chain.  We
have also verified that the resulting distributions show excellent
agreement across all five chains.  For example, the Gelman-Rubin test
statistic, $\hat{R}$, approaches 1
from above as a Markov chain approaches convergence.  While sets of
Markov chains with $\hat{R}$ less that 1.1, or even 1.2, are frequently
used for inference, for the Markov chains used in this analysis,
$\hat{R}$ was less than 1.001 for each fit parameter and auxiliary variable,
and for the predictions of the radial velocity at 40 future epochs.
 Thus, we are very confident in the reliability of our Markov chains.

We use the results from our Markov chains as the basis for our
dynamical analyzes.  When considering non-coplanar cases, we
independently draw the inclinations from isotropic distributions and
the longitude of ascending node from a uniform distribution.  For each
state in the Markov chain, we calculate the planet masses and
semi-major axes iteratively and treat them as Jacobi elements.
Finally, we have performed direct $N$-body integrations of systems
sampled from our Markov chains. For these integrations, we
chose a variety of initial epochs varying by several times the orbital
period of planet d.  By comparing the $\chi2$ of the fit determined by
the analytic and $N$-body models, we have determined that our results
are not affected by including the effects of mutual planetary
perturbations in the fitting procedure. \\

\bigskip

\hrule
\medskip

\noindent {\bf Additional References for Supplementary Methods}

\begin{enumerate}
\item[s1] Gelman, A., Carlin, J.B., Stern, H.S. \& Rubin, D.B.
 {\it Bayesian Data Analysis\/} (Chapman \& Hall/CRC, New York, 2003).
\item[s2] Ford, E.B. Quantifying the uncertainty in the orbits of
extrasolar planets.
  {\it Astron.\ J.\/}, in press (2005).
\end{enumerate}

\clearpage

\begin{center}
\noindent {\bf Supplementary Discussion}
\end{center}

\noindent {\bf Choice of Initial Conditions for Figure 4 }

We have adopted this simple two--planet configuration for computational
convenience. In this configuration (two planets on circular orbits
perturbing each other significantly, with the other planets distant
enough to be negligible), the dynamical stability limit is known
analytically and is sharply defined$^{21}$. Therefore we place
the two planets very near that limit in our initial condition. As
discussed in previous papers, a variety of scenarios would lead
naturally to two planets approaching this limit$^{8,17,18}$. For
example, the outer planet might be migrating inward through coupling
with an outer gaseous disk. Once the stability limit is reached, the
system evolves quickly (on the orbital timescale) until the strong
scattering occurs and one planet is ejected, while the gas becomes
irrelevant (as the viscous timescale is much longer).

In addition, the chaotic evolution of a multi--planet system (with
more than two planets perturbing each other significantly) can also
easily lead to strong scattering, sometimes after a long period of
seemingly ``stable'' evolution$^{20,21}$.  Although equally plausible,
this is less convenient computationally since, with more than two
planets, the stability limit is not known analytically and not sharply
defined, so numerical experimentation would be needed to find a case
that could produce a final state resembling the current $\upsilon$ And
configuration, possibly after a very long dynamical integration.

With more than two planets, the timescale for growth of the
instability and the occurrence of a strong scattering involving
$\upsilon$ And~d can be arbitrarily long$^{21,s3}$ and could easily
exceed the $\sim\!10^7$ yr timescale on which the gas is expected to
be lost from the protoplanetary disk.  If gas were still present when
the scattering occurs, the final outcome would be modified, but only
the details of the dynamical evolution would change. For example, if
gas drag produces some damping of the eccentricity, then a slightly
stronger scattering may be needed to produce the same final
eccentricity for the retained planet.

Finally, we want to emphasize that, while reproducing the exact
parameters of any particular observed system always requires ``fine
tuning'' in an obvious sense, what is important here is that our
mechanism can --- naturally and without any fine tuning --- provide
the very short timescale required for the initial perturbation that
left $\upsilon$ And~d on an eccentric orbit while leaving $\upsilon$ And~c on a
perfectly circular orbit. \\

\bigskip

\noindent {\bf Perturbations Due to a Gas Disk }

The possibility that the impulsive perturbation to $\upsilon$ And~d
was delivered by a massive exterior gaseous disk with a large
viscosity was mentioned in a previous study$^{4}$.  However, we can
show that this possibility is now firmly excluded for this system.
If the eccentricity of $\upsilon$ And d had been induced by an outer
disk, the eccentricity growth time would have to be considerably
shorter than the secular timescale.  In addition, after the
eccentricity grew to $\sim\!0.3$, the perturbation must stop
suddenly. Otherwise the eccentricity of the middle planet, $\upsilon$
And~c, would also start growing and its ``initial'' value would no
longer be compatible with the minimum we derive in the secular
solution.

Quantitatively, this is a very stringent requirement. We have
performed additional numerical integrations for the outer two planets,
starting with their current masses and orbital periods, but on
circular orbits. In addition to the mutual gravitational
perturbations, we include a simple semi--analytic model of angular
momentum dissipation acting on $\upsilon$ And~d only.  The dissipation
rate ($\dot{J}$) is constant for a time $\Delta t$ and then disappears
(completely and instantaneously). We have performed multiple
simulations holding the product $\dot{J} \Delta t$ fixed to reproduce
the current eccentricity of $\upsilon$ And~d. If we impose the constraint
that the eccentricity of $\upsilon$ And~c must remain $\lo0.01$ (the
current best--fit value of the minimum is 0.005) after $\Delta t$,
then this model provides an upper limit of $\Delta t \lo 100$ years.
Thus, it is not sufficient to impose an eccentricity growth
time shorter than the secular period of $\sim\!10^4$ years. 

A timescale for eccentricity growth by viscous coupling to a disk as
short as $\lo100$ years would require both an implausibly massive disk
and a very high effective viscosity.  The possibility of eccentricity
growth caused by interactions with a disk is rather
controversial$^{s4}$, particularly for planets less massive than
$10-20\,M_{\rm J}$.  Nevertheless, we have estimated the timescale for
eccentricity using the only detailed theory of eccentricity excitation
by viscous coupling to a disk in the astrophysical literature$^{s5}$.
We find that a timescale for eccentricity growth as short as
$\sim\!100$ years would require a mass in the relevant part of the
disk, $\Sigma r^2 \sim\!40\,M_J$, ten times more than the mass of
the planet, even with an implausibly large disk viscosity parameter,
$\alpha \sim\!0.1$.
Instead, using more typical parameters$^{s5}$ (for a $1\, M_J$ planet
at 1 AU, with disk parameters $r/h=25$, $\alpha=0.001$, and $w/r =
0.5$), it would require $\sim\!10^7$ years for the eccentricity to
grow from 0.01 to 0.3.  Eccentricity growth on this very long
timescale would instead lead to apsidal resonance between $\upsilon$
And~c and~d with a small amplitude libration$^{10}$, which we have
ruled out in the main text.  In addition, neither this theory$^{s5}$
nor any other published theory of eccentricity excitation due to a gas
disk provides a mechanism for making the perturbation cease quickly
($\lo100$ years) after a phase of very rapid eccentricity growth.

What is truly unique about planet--planet scattering is that it
provides {\em both\/} a very short timescale for eccentricity growth (the
dynamical timescale on which the instability develops) {\em and\/} the same
very short timescale for removing the perturbation (since the extra
planet gets ejected as a result of the scattering). \\

\bigskip

\hrule
\medskip

\noindent {\bf Additional References for Supplementary Discussion}

\begin{enumerate}
\item[s3] Chambers, J.E., Wetherill, G.W. \& Boss, A.P. 
The Stability of Multi-Planet Systems.
 {\it Icarus \/} {\bf 119}, 261--268 (1996).
\item[s4] Papalouizou, J.C.B., Nelson, R.P. \& Masset, F. 
Orbital eccentricity growth through disc-companion tidal interaction.
 {\it Astron. \& Astrophys. \/} {\bf 366}, 263--275 (2001).
\item[s5] Goldreich, P. \& Sari, R.
Eccentricity Evolution for Planets in Gaseous Disks.
 {\it Astrophys.\ J.\/} {\bf 585}, 1024--1037 (2003).

\end{enumerate}

\end{document}